\documentclass[
]{jpsj2}
\usepackage{amssymb}
\usepackage{latexsym}
\usepackage{graphicx}
\usepackage{amsmath}

\newcommand{\ra}{\rangle}
\newcommand{\la}{\langle}
\newcommand{\ua}{\uparrow}
\newcommand{\da}{\downarrow}
\newcommand{\Sin}{{\rm sin}}
\newcommand{\Cos}{{\rm cos}}

\newcommand{\pa}{\psi_{\alpha\sigma}}
\newcommand{\pad}{\psi^{\dagger}_{\alpha\sigma}}
\newcommand{\pb}{\psi_{\beta\sigma'}}
\newcommand{\dx}{\frac{\partial}{\partial x}}
\newcommand{\fint}{\int^{\infty}_{-\infty}}
\newcommand{\psd}{\psi^{\dagger}}

\def\runauthor{Online-Journal Subcommittee}

\title{Nonequilibrium Kondo problem with spin-dependent chemical potentials: Exact results
 \textit{}
}

\author{Hosho Katsura
\thanks{E-mail address: katsura@appi.t.u-tokyo.ac.jp}
}

\inst{%
Department of Applied Physics, University of Tokyo, Hongo, Bunkyo-ku, Tokyo 113-8656, Japan
}

\recdate{\today}

\abst{%
We give a brief review of the Kondo effect and exactly solve 
the nonequilibrium Kondo problem at the special point in the parameter 
space of the model with spin-dependent chemical potentials. 
Using the obtained solution, we compute several experimentally observable 
quantities: charge current, spin current and magnetic properties. 
Applications to the physically important cases such as the case where 
the impurity spin is placed in between two ferromagnetically polarized 
leads are discussed.
}

\kword{%
kondo problem, spin current, nonequilibrium steady state, bosonization
}

\begin{document}
\maketitle

\section{Introduction}
In most metals, the resistivity monotonically decreases with the decrease of temperature because it is dominated by phonon scattering. On the other hand, the metals with magnetic impurities such as Fe have the resistance minimum at a certain temperature, known as Kondo temperature $T_K$, and under $T_K$ the resistivity approaches unitary limit as the temperature approaches absolute zero. The experimental discovery of this effect dates back to the early 30s. The reason for the resistance minimum was not known at that time. In the 60s, however, a very significant advance of the theory for the effect was developed by Kondo \cite{Kondo}. Kondo's calculation of the resistivity, which was to explain this minimum, was based on the $s$-$d$ model where the local impurity spin $S$ is coupled via an exchange interaction $J$ with the conduction electrons of the host metal. Using third order perurbation theory in coupling $J$, he showed that this $s$-$d$ interaction leads to singular scattering of the conduction electrons near the Fermi surface and a ${\rm ln}T$ contribution to the resistivity. The ${\rm ln}T$ term increases at low temperatures for an antiferromagnetic coupling and when this term is included with the phonon contribution to the resistivity, i.e., $T^5$ term, it is sufficient to explain the observed resistance minimum. Since his perturbation calculations could not be valid at low temperatures, a more comprehensive theory was needed to explain the low temperature behavior. This task attracted a lot of theoretical interests in the late 60s and early 70s. More extensive perturbative calculations which sum the most divergent terms were performed by Abrikosov \cite{Ab}. The scaling idea introduced by Anderson provided a new theoretical framework for the Kondo problem \cite{Anderson}. This idea was taken over by Wilson's renormalization group method discussed below. Another perturbative approach was developed in the study of the Anderson model, which is more microscopic than $s$-$d$ model, where perturbative calculations for Coulomb repulsion $U$ have no divergence \cite{Yamada}. Nonperturbative approaches to the Kondo problem will be overviewed below. From the present point of view, the Kondo effect is understood as a precursory phenomenon of the formation of the bound state between the conduction electrons and the impurity spin, known as {\it Kondo singlet}. The theoretical structure underlying the Kondo problem is similar to the one in Quantum Chromodynamics (QCD) and the Kondo effect corresponds to the confinement of quarks and gluons.\\
Recently, the Kondo effect is revived in the context of mesoscopic physics such as semiconductor quantum dot devices where the quantum dot spin and the electrons in electrodes corresnpond to the impurity spin and the conduction electrons in the original Kondo problem, respectively \cite{Tarucha}. The crucial difference between the usual Kondo effect and that in mesoscopic systems is the nonequilibrium effect, i.e., the voltage vias between electrodes. It is very difficult to approach theoretically to the nonequilibrium situation of the Kondo problem. One of the standard method to treat nonequilibrium problems perturbatively is the Keldysh Green's function method (discussed in ref. \ref{Oguri} in connection with the Kondo effect). Nonperturbative approaches will be discussed in detail below.\\ 
Before discussing the nonequilibrium Kondo problem, let us briefly overview the nonperturbative approaches to the equilibrium Kondo problem (more detailed information can be found in ref. \ref{Hewson} and \ref{Shiba}). The first important contribution developed by Wilson in the 70s is the Numerical Renormalization Group (NRG) method \cite{Wilson} which provides a way to numerically perform scale transformations proposed by Anderson. By using NRG, Wilson has obtained definitive results for the ground state and low temperature behavior for the spin half $s$-$d$ model. The second development in the late 70s and early 80s is the discovery of exact solutions to the Kondo problem. By using Bethe Ansatz (BA) method which is a powerful tool to solve one-dimensional integrable models such as the Heisenberg model and the Hubbard model,  Andrei \cite{Andrei} and Wiegmann \cite{Wiegmann} solved the spin half $s$-$d$ model independently of each other. The critical idea, already used in Wilson's NRG, in BA approach is the mapping to the equivalent one-dimensional model. After these works, Wiegmann showed that BA approach could be extended to give exact results for the Anderson model \cite{Wiegmann2}. The most recent nonperturbative approaches by Affleck and Ludwig in the 90s are based on Conformal Field Theory (CFT) \cite{Affleck}. They have formulated the Kondo problem as a boundary quantum critical phenomenon and have applied CFT to it. First, they found the nontrivial strong-coupling fixed point where the absorption of the impurity spin is elegantly described by the deformation of Kac-Moody algebra, infinite dimensional Lie algebra, for the spin part. Second, they assumed that the absorption process at the point away from the fixed point is governed by Kac-Moody fusion rules, the analogue of addition of angular momentum in CFT, and extracted the anomalous non-Fermi liquid(NFL) behavior of the multichannel Kondo problem introduced by A. Blandin and P. Nozi$\acute{\rm e}$res. After the works of Affleck and Ludwig, Emery and Kivelson have mapped the two-channel Kondo problem into the resonant-level model, which is equivalent to noninteracting fermions for a particular value of the $z$ component of the exchange similar to the Toulouse limit\cite{Toulouse}, by using the standard Abelian bosonization technique \cite{Emery}. Since the resultant resonant-level model is a quadratic form in electrons and impurity spin operators at this special Fermi Liquid (FL) fixed point, the absorption of the impurity spin is described by the unitary transformation which diagonalizes the above-mentioned quadratic form. 

We shall now leave the equilibrium case and turn to the nonperturbative approaches to the nonequilibrium Kondo problem. The first contribution is given by Schiller and Hershfield \cite{Schiller,Schiller2}. They combined the $Y$-operator method \cite{Hershfield} developed by Hershfield with the Emery-Kivelson solution discussed above and then obtained the exact solution for the nonequilibrium Kondo problem at a special point in the parameter space. After this work, Delft {\it et al} have developed the CFT approach combined with Hershfield's $Y$-operator formulation \cite{Delft}. Recently, Mehta and Andrei have applied the Nonequilibrium Bethe Ansatz (NEBA) to the computation for steady state properties of quantum impurity problems such as interacting resonant level model(IRLM) \cite{Mehta}. Here we should note that the comprehensive nonperturbative understanding of the nonequilibrium Kondo problem is still an open problem and requires more extensive studies.

In this article, we extend the Schiller-Hershfield solution to the nonequilibrium Kondo problem to include the case where the chemical potential of each lead depends on the spin of conduction electrons. Actually, these situations are often discussed in the area such as {\it spintronics}. It is different from Ref.\ref{Schiller} and \ref{Schiller2} that we employ a description of nonequilibrium steady states (NESS) \cite{Tasaki, Takahashi}, developed in the context of mathematically rigorous theory such as ${\rm C}^*$-algebra, instead of using Hershfield's $Y$-operator. The obtained results given below are generalization of those in ref. \ref{Schiller} and \ref{Schiller2} and reproduce their results as limiting cases.

\section{Model and mapping}
\begin{figure}
\includegraphics[width=14cm,clip]{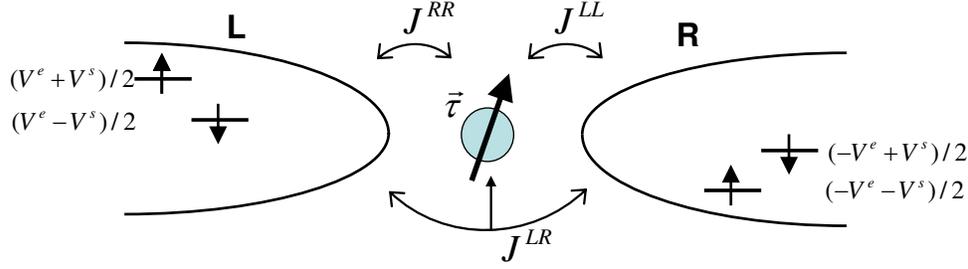}
\caption{Schematic description of our system. A tunnel junction consists of 
two leads of noninteracting spin-1/2 electrons and a spin-1/2 impurity spin 
placed in between the two leads. The energy levels denote the spin-dependent chemical potentials of the conduction electrons.}
\label{leads}
\end{figure}

The physical system under consideration is shown schematically in Fig.(\ref{leads}). It consists of left(L) and right(R) leads of noninteracting spin-1/2 electrons, which interact via an exchange coupling with a spin-1/2 impurity spin placed in between the two leads. Following Ref.\cite{Affleck}, the conduction-electron channels that couple to the impurity are reduced to one-dimensional fields $\psi_{\alpha\sigma}(x)$, where $\alpha = L/R$ and $\sigma =\ua,\da$ are the lead and spin indices respectively. In what follows, we assume that the conduction-electron dispersion around the Fermi level is linear, i.e., $\epsilon_k = \hbar v_F k$, where $\epsilon_k$ and $k$ are measured relative to the Fermi level and Fermi wave number respectively. The most general form of the $s$-$d$ Hamiltonian $H$ is given by
\begin{eqnarray}
H =i\hbar v_F \sum_{\alpha=L,R} \sum_{\sigma=\ua,\da} \int^{\infty}_{^\infty} 
\pad(x)\dx \pa(x) dx \nonumber \\
+\sum_{\alpha,\beta=L,R} \sum_{\lambda=x,y,z} J^{\alpha\beta}_{\lambda}s^{\lambda}_{\alpha\beta}\tau^{\lambda} -\mu_B g_i H \tau^z,
\label{Ham}
\end{eqnarray}
where $\vec \tau$ denotes the impurity spin and $s^{\lambda}_{\alpha\beta}$ is defined as
\begin{equation}
s^{\lambda}_{\alpha\beta} \equiv \sum_{\sigma,\sigma'}\pad(0) \frac{\sigma^{\lambda}_{\sigma\sigma'}}{2} \pb(0),
\end{equation}
where $\sigma^{\lambda},~(\lambda=x,y,z)$ are a Pauli matrices. And $\mu_B$ and $g_i$ are the Bohr magneton and the impurity Land$\acute{\rm e}$ $g$ factor, respectively. In addition to the above Hamiltonian $H$, we consider the following nonequilibrium conditions:
\begin{eqnarray}
Y_0 &=&Y^e_0 + Y^s_0, \nonumber
\\
Y^e_0 &=& \frac{eV^e}{2}\sum_{\sigma}\fint[\psi^{\dagger}_{L\sigma}\psi_{L\sigma}-\psi^{\dagger}_{R\sigma}\psi_{R\sigma}]dx, \nonumber \\
Y^s_0 &=& \frac{eV^s}{2}\sum_{\sigma}[\sigma\psi^{\dagger}_{L\sigma}\psi_{L\sigma}-\sigma\psi^{\dagger}_{R\sigma}\psi_{R\sigma}]dx,
\label{necond}
\end{eqnarray}
where $V^e$ and $V^s$ correspond to a usual electronic voltage and a spin voltage respectively. In Ref. \ref{Schiller} and \ref{Schiller2}, Schiller and Hershfield have considered the case where only $V_e$ exists. On the other hand, there are some situations, such as ferromagnetic leads cases, where the spin voltage $V_s$ exists and plays a key role in the spin transport phenomena such as {\it spincurrent}. Therefore, the nonequilibrium Kondo problem with the above more general nonequilibrium conditions needs to be newly formulated. \\
We shall discuss how to diagonalize the above Hamiltonian. First, we restrict the region of the $J^{\alpha\beta}_{\lambda}$ parameter space to the case where $J^{\alpha\alpha}_x=J^{\alpha\alpha}_y \equiv J^{\alpha}_{\perp}$, $J^{LR}_x~J^{LR}_y \equiv J^{LR}_{\perp}$, $J^{LR}_z=J^{RL}_z=0$, and $J^{LL}_z=J^{RR}_z \equiv J_z$ takes a particular value (discussed below). We note that setting $J^{LR}_z=0$ produces no significant difference on the low-energy physics since, in the renormalization group sense, $J^{LR}_z \ne 0$ is generated on scaling from weak coupling \cite{Furusaki}. Next, we bosonize eq.(\ref{Ham}) and eq.(\ref{necond}) respectively by following the Emery-Kivelson solution of the two-channel Kondo problem \cite{Emery}. 
We introduce four different boson fields to account for four different fermion fields as
\begin{eqnarray}
&&\psi_{L \ua}(x)
={\frac{1}{\sqrt{2 \pi a}}} (-1)^{N_{L \da}+N_{R \ua}+N_{R \da}} e^{-i\Phi_{L \ua}(x)},
\\
&&\psi_{L \da}(x)
={\frac{1}{\sqrt{2 \pi a}}}(-1)^{N_{R \ua}+N_{R \da}}e^{-i\Phi_{L \da}(x)},
\\
&&\psi_{R \ua}(x)
={\frac{1}{\sqrt{2 \pi a}}}(-1)^{N_{R \da}}e^{-i\Phi_{R \ua}(x)},
\\
&&\psi_{R \da}(x)
={\frac{1}{\sqrt{2 \pi a}}}e^{-i\Phi_{R \da}(x)},
\label{bosonize}
\end{eqnarray}
where $a$ is a high-energy cut-off(lattice spacing), and number operators for fermion $N_{i,\sigma}~(i=L,R,\sigma=\ua,\da)$ are defined as
\begin{equation}
N_{i \sigma}
=\int_{-\infty}^{\infty}dx :\psi^{\dagger}_{i\sigma}(x) \psi_{i\sigma}(x):
=\frac{1}{2 \pi} \int_{-\infty}^{\infty}dx \frac{\partial \Phi_{i\sigma}(x)}{\partial x},
\end{equation}
and satisfy the canonical commutation relations $[\Phi_{i\sigma}(x), N_{j\sigma}]=i \delta_{ij}\delta_{\sigma\sigma'}$. The boson fields $\Phi_{i,\sigma}$ also satisfy standarad commutation relations 
\begin{equation}
[\Phi_{i\sigma}(x), \Phi_{j\sigma'}(y)]=-i \pi \delta_{ij}\delta_{\sigma\sigma'} {\rm sgn}(x-y),
\label{boc}
\end{equation}
and hence we can make sure of the following correspondence between the fermion fields and boson fields:
\begin{align}
2\pi :\psi^{\dagger}_{i\sigma}(x) \psi_{i\sigma}(x):&
=\frac{\partial \Phi_{i\sigma}(x)}{\partial x}
\\
i v_F \int_{-\infty}^{\infty}dx:\psi^{\dagger}_{i\sigma}(x){\frac{\partial}{\partial x}}\psi_{i\sigma}(x):&
=\frac{v_F}{4 \pi}\int_{-\infty}^{\infty}dx:({\frac{\partial \Phi_{i\sigma}(x)}{\partial x}})^2:
\end{align}
where $:AB...:$ denotes the normal ordering of operator products $AB...$. 
Furthermore, we introduce four new boson fields constructed by the above four boson fields in eq.(\ref{bosonize}) as
\begin{align}
\Phi_c&=\frac{1}{2}(\Phi_{L \ua} + \Phi_{L \da} +\Phi_{R\ua}+\Phi_{R\da}),\nonumber
\\
\Phi_s&=\frac{1}{2}(\Phi_{L \ua} - \Phi_{L \da} +\Phi_{R\ua}-\Phi_{R\da}),\nonumber
\\
\Phi_f&=\frac{1}{2}(\Phi_{L \ua} + \Phi_{L \da} -\Phi_{R\ua}-\Phi_{R\da}),\nonumber
\\
\Phi_{sf}&=\frac{1}{2}(\Phi_{L \ua} - \Phi_{L \da} -\Phi_{R\ua}+\Phi_{R\da}),
\label{csf}
\end{align}
where subscripts $c, s, f, sf$ correspond to collective charge, spin, flavor (left minus right), and spin-flavor modes respectively. 
Now, the Hamiltonian and nonequilibrium conditions can be rewritten in terms of the boson fields in eq.(\ref{csf}) as
\begin{align}
H &=\frac{\hbar v_F}{4\pi} \sum_{\nu =c,s,f,sf} \int^{\infty}_{-\infty}dx \left(\frac{\partial \Phi_{\nu}}{\partial x} \right)^2 
\nonumber \\
&+\frac{J^+}{\pi a}[ -\tau^x \Sin(\chi_s) +\tau^y \Cos(\chi_s)] \Cos(\chi_{sf})
\nonumber \\
&-\frac{J^-}{\pi a}[\tau^x \Cos(\chi_s) +\tau^y \Sin(\chi_s)] \Sin(\chi_{sf})
\nonumber \\
&-\frac{J^{LR}_{\perp}}{\pi a}[\tau^x \Cos(\chi_s)+\tau^y \Sin(\chi_s)]\Sin(\chi_{f})
\nonumber \\
&+\frac{J_z}{2\pi}\tau^z \frac{\partial \Phi_s}{\partial x}|_{x=0} -\mu_B g_i H \tau^z,
\end{align}
and
\begin{align}
Y^e &=\frac{eV^e}{2\pi}\fint dx \frac{\partial \Phi_f(x)}{\partial x},
\nonumber \\
Y^s &=\frac{eV^s}{2\pi}\fint dx \frac{\partial \Phi_{sf} (x)}{\partial x},
\end{align}
where $J^{\pm}$ are the even and odd combinations of $J^{LL}_{\perp}$ and $J^{RR}_{\perp}$, i.e., $J^{\pm} \equiv (J^{LL}_{\perp}+J^{RR}_{\perp})/2$
and $\chi_s, \chi_f$ and $\chi_{sf}$ are defined as
\begin{align}
\chi_s &\equiv\Phi_s(0) -\frac{\pi}{2}(N_c-N_s)+\frac{\pi}{2},
\nonumber \\
\chi_f &\equiv\Phi_f(0) - \frac{\pi}{2}(2N_c -N_f -N_{sf}),
\nonumber \\
\chi_{sf} &\equiv\Phi_{sf}(0) -\frac{\pi}{2}(N_f -N_{sf}).
\end{align}
The next step is to employ the canonical transformation $H' =UHU^{\dagger}$ and $Y' = UY_0U^{\dagger}$ with $U={\rm exp}(i\chi_s \tau^z)$.
Since $\chi_s$ commutes with both $\chi_f$ and $\chi_{sf}$, the transformed Hamiltonian $H'$ is written as
\begin{align}
H' &=\frac{\hbar v_F}{4\pi} \sum_{\nu =c,s,f,sf}\fint dx \left(\frac{\partial \Phi_\nu(x)}{\partial x} \right)^2 +\frac{J^+}{\pi a} \tau^y \Cos(\chi_{sf}) - \frac{J^-}{\pi a}\tau^x \Sin(\chi_{sf})
\nonumber \\
&-\frac{J^{LR}_{\perp}}{\pi a}\tau^x \Sin(\chi_f) +\left[\frac{J^z}{2\pi}-\hbar v_F\right]
\frac{\partial \Phi_s(0)}{\partial x}\tau^z -\mu_B g_i H \tau^z,
\end{align}
where we have used the commutation relation
\begin{equation}
\left[e^{i\chi_s \tau^z}, \frac{\partial \Phi_{\nu}(x)}{\partial_x}\right] =-2\pi i \delta_{\nu s}\delta(x) e^{i\chi_s \tau^z},
\end{equation}
and have neglected divergent constant (the validity has been discussed in ref. \ref{AL}).
The nonequilibrium conditions $Y^e_0$ and $Y^s_0$ are unaffected by the above canonical transformation, i.e., $Y' =Y_0$.
The last step is {\it refermionization} of the above bosonized operators.
To this end, we first express the spin $\vec \tau$ in terms of fermion operator as $d = i\tau^+$.
The four boson fields $\Phi_{\nu}~(\nu =c,s,f,sf)$ are refermionized according to
\begin{align}
\psi_f(x)&=\frac{e^{i\pi d^{\dagger}d}}{\sqrt{2\pi a}} e^{-i(\Phi_f(x)-\frac{\pi}{2}(2N_c-N_f-N_{sf}))},
\nonumber \\
\psi_{sf}(x)&=\frac{e^{i\pi d^{\dagger}d}}{\sqrt{2\pi a}} e^{-i(\Phi_{sf}(x)-\frac{\pi}{2}(N_f-N_{sf}))},
\end{align}
etc.
Here the phase $i\pi d^{\dagger}d$, similar to the Jordan-Wigner transformation, guarantees the anticommutation relations between $d$ and $\psi_{\nu}(x)$. \\
Once these steps are completed, we can obtain the refermionized forms for $H'$ and $Y'$:
\begin{align}
H' &=i\hbar v_F \sum_{c,s,f,sf} \fint dx \psi^{\dagger}_{\nu}(x) \dx \psi_{\nu}(x)
+\frac{J^+}{2\sqrt{2\pi a}}[\psi^{\dagger}_{sf}(0)+\psi_{sf}(0)](d^{\dagger}-d)
\nonumber \\
&+\frac{J^{LR}_{\perp}}{2\sqrt{2\pi a}}[\psi^{\dagger}_{f}(0)-\psi_{f}(0)](d^{\dagger}+d)
+\frac{J^-}{2\sqrt{2\pi a}}[\psi^{\dagger}_{sf}(0)-\psi_{sf}(0)](d^{\dagger}+d)
\nonumber \\
&+[\mu_B g_i H -(J_z-2\pi \hbar v_F):\psi^{\dagger}_s(0)\psi_s(0):]\left(d^{\dagger}d-\frac{1}{2}\right),
\label{pd}
\end{align}
and
\begin{align}
(Y^e_0)' &= eV^e \fint dx \psi^{\dagger}_f(x) \psi_f(x),
\nonumber \\
(Y^s_0)' &= eV^s \fint dx \psi^{\dagger}_{sf}(x) \psi_{sf}(x).
\end{align}
Here we can immediately notice that (\ref{pd}) would be a quadratic form in fermion operators if we set $J_z=2\pi \hbar v_F$. Since the nonequilibrium condition $Y'$ is originally a quadratic form, both $H'$ and $Y'$ reduce to quadratic form at this special point. Hence the strongly interacting nonequilibrium problem maps to a noninteracting one.
For convenience, we introduce new {\it Majorana} fermions as
\begin{equation}
{\hat a} = \frac{d+d^{\dagger}}{\sqrt{2}},\hspace{1cm}{\hat b} = \frac{d^{\dagger}-d}{i\sqrt{2}}.
\end{equation}
Here we note that Majorana fermions satify ${\hat a}^2={\hat b}^2=1/2$.
After Fourier transformation, $H'$ and $Y$ for $J_z=2\pi \hbar v_F$ can be rewritten as
\begin{align}
H' &=\sum_{\nu=f,sf}\sum_k \epsilon_k \psi^{\dagger}_{\nu,k}\psi_{\nu,k}
-i \mu_B g_i H {\hat a}{\hat b}
+i\frac{J^+}{2\sqrt{\pi a L}} \sum_k (\psi^{\dagger}_{sf,k}+\psi_{sf,k})\hat b
\nonumber \\
&+\frac{J^{LR}_{\perp}}{2\sqrt{\pi a L}} \sum_k (\psi^{\dagger}_{f,k}-\psi_{f,k})\hat a
+\frac{J^-}{2\sqrt{\pi a L}}\sum_k (\psi^{\dagger}_{sf,k}-\psi_{sf,k})\hat a,
\label{pdk}
\end{align}
\begin{equation}
\begin{array}{l}
Y'= eV^e \sum_k \psd_{f,k}\psi_{f,k} + eV^s \sum_k \psd_{sf,k}\psi_{sf,k},
\end{array}
\label{Y'}
\end{equation}
where $L$ is the size of the system, $\epsilon_k=\hbar v_F k$ and we have neglected the terms involve the $\psi_c$ and $\psi_s$ which are decoupled from $\hat a$ and $\hat b$.
\\
\section{Solution and observables}
In this section, we shall show how to diagonalize eqs.(\ref{pdk}) and (\ref{Y'}) simultaneously. After obtaining the exact solution, we will calculate experimentally observable quantities such as charge current, spin current and magnetic properties. \\
Since the precise procedure for diagonalization of $H'$  is given in ref. \ref{Schiller2}, we shall only outline the main steps briefly. The first step is to construct scattering states $c^{\dagger}_{\nu,k}$ which diagonalize $H'$ and are defined by the operator equation
\begin{equation}
[c^{\dagger}_{\nu,k}, H']=-\epsilon_k c^{\dagger}_{\nu,k} +i\eta(\psd_{\nu,k}-c^{\dagger}_{\nu,k}) \hspace{1cm} (\nu=f,sf),
\end{equation}
where the positive infinitesimal $\eta$ is introduced to guarantee appropriate boundary conditions. Detailed derivation of the scattering-states operators are given in the APPENDIX B in ref. \ref{Schiller2}. Using $c_{\nu,k}$, our Hamiltonian $H'$ can be written in a diagonal form as
\begin{equation}
H'= \sum_{\nu=f,sf} \sum_k \epsilon_k  c^{\dagger}_{\nu,k}c_{\nu,k}
\end{equation}
Next we need to expand original operators, $\psi^{(\dagger)}_{\nu,k}, \hat a$ and $\hat b$, in terms of scattering-states operators $c^{(\dagger)}_{\nu,k}$. TABLE II in ref. \ref{Schiller2} summarizes the results of expansion. At this point, we employ a description of nonequilibrium steady states (NESS) instead of the $Y$-operator method developed by Hershfield. In NESS approach, once we can obtain steady state density matrix $\rho_+$, we impose the following conditions on the two-point function:
\begin{equation}
\la c^{\dagger}_{\mu,k}c_{\nu,k'}\ra = F_{\mu}(\epsilon_k)\delta_{\mu \nu}\delta_{k k'} \hspace{1cm} (\mu, \nu=f,sf),
\label{impose}
\end{equation} 
where $\la ... \ra \equiv {\rm Tr}(...\rho_+)$ stands for the NESS average and $F_{\mu}(\epsilon)$ are defined by the ordinary Fermi distribution function $f(\epsilon)$ as $F_f(\epsilon) \equiv f(\epsilon-eV^e)$ and $F_{sf}(\epsilon) \equiv f(\epsilon-eV^s)$ respectively. We should note here that there is a somewhat subtle point to use $C^*$-algebraic approach, since we involved non-local operators such as $N_i$ in our bosonization procedure.

Now, we are ready to calculate the observable quantities. The physical observables of interest such as charge current $I_c$, spin current $I_s$ and impurity magnetization $M^z$ are rewritten after the canonical transformation $U$:
\begin{align}
I'_c &=i\frac{eJ^{LR}_{\perp}}{2 \hbar \sqrt{\pi a L}}\sum_k (\psd_{f,k}+\psi_{f,k})\hat a,
\\
I'_s &=i\frac{J^-}{2 \hbar \sqrt{\pi a L}}\sum_k (\psd_{sf,k}+\psi_{sf,k}){\hat a} -\frac{J^+}{2 \hbar \sqrt{\pi a L}}\sum_k (\psd_{sf,k}-\psi_{sf,k}){\hat b},
\\
M^z &=i \mu_B g_i {\hat a}{\hat b},
\end{align}
(the derivation can be found in ref. \ref{Schiller2}). From now on, we focus on the average of transformed operators, since $\la ...\ra ={\rm lim}_{t \to +\infty}{\rm Tr}(...e^{-iHt}\rho_0 e^{iHt}) = {\rm lim}_{t\to +\infty}{\rm Tr}(U...U^{\dagger}\cdot Ue^{-iHt}U^{\dagger}\cdot U \rho_0 U^{\dagger} \cdot U e^{iHt} U^{\dagger}) = {\rm lim}_{t \to +\infty}{\rm Tr}(U...U^{\dagger} e^{-iH't}\rho'_0 e^{iH't}) =\la U...U^{\dagger} \ra$, where $\rho_0$ is the density matrix for the initial state and we have used the cyclic property of the {\it trace}. Using the condition for the two-point correlation function (\ref{impose}), we can obtain the average of $I'_c, I'_s$ and $M^z$ due to the quadratic nature of them. (Generally speaking, we can also compute higher-order terms by using Wick's theorem for $c_{\nu,k}$).\\
\\
\noindent
{\it Charge Current and Spin Current}\\
First, let us discuss the charge current and spin current in our system. Since the charge current obtained here is exactly equal to the result in \cite{Schiller2}, we only write down the result:
\begin{equation}
I_c(V)=\la I'_c(V)\ra = \frac{e \Gamma_1}{2 \pi \hbar} \fint d\epsilon A_a(\epsilon)[f(\epsilon-eV^e)-f(\epsilon+eV^e)],
\label{Ic}
\end{equation}
where 
\begin{equation}
A_a(\epsilon)=-{\rm Im}\left\{\frac{\epsilon+i\Gamma_b}{(\epsilon+i\Gamma_a)(\epsilon+i\Gamma_b)-(\mu_B g_i H)^2} \right\},
\end{equation}
and $\Gamma_{\alpha},(\alpha=1,a,b)$ are tabularized below.
Here, we note that $I_c(V)$ depends only on $V^e$ and is independent of the spin voltage $V^s$. \\
\\
\begin{table}
\begin{tabular}{cc}
\hline
Symbol & Definition\\
\hline
$\Gamma_a$ & $[(J^{LR}_{\perp})^2+(J^-)^2]/4\pi a \hbar v_F $\\
$\Gamma_b$ & $(J^+)^2/4\pi a \hbar v_F$\\
$\Gamma_1$ & $(J^{LR}_{\perp})^2/4\pi a \hbar v_F$\\
$\Gamma_2$ & $(J^-)^2/4\pi a \hbar v_F$\\
$\Gamma_L$ & $(J^{LL}_{\perp})^2/4\pi a \hbar v_F$\\
$\Gamma_R$ & $(J^{RR}_{\perp})^2/4\pi a \hbar v_F$\\
$\Gamma_m$ & $J^{LR}_{\perp}J^-/4\pi a \hbar v_F$\\
$\Gamma_p$ & $J^{LR}_{\perp}J^+/4\pi a \hbar v_F$\\
\hline
\end{tabular}
\caption{Symbols and their definitions.}
\end{table}
\\
\\
We will now leave the discussion of charge current and turn to that of spin current. We obtain the following result which includes the one 
in ref. \ref{Schiller2} as a specific case:
\begin{align}
I_s(V)&=\la I'_s \ra 
\nonumber \\
&=\frac{1}{2 \pi \hbar} \fint d\epsilon \left[\frac{f(\epsilon-eV^e)+f(\epsilon+eV^e)}{2}-\frac{f(\epsilon-eV^s)+f(\epsilon+eV^s)}{2} \right]
\nonumber \\
&\times \frac{(\mu_B g_i H)(\Gamma_L-\Gamma_R)\Gamma_1\epsilon}{|(\epsilon+i\Gamma_a)(\epsilon+i\Gamma_b)-(\mu_B g_i H)^2|^2}
\nonumber \\
&+\frac{1}{2\pi \hbar} \fint d\epsilon [f(\epsilon-eV^s)-f(\epsilon+eV^s)][\Gamma_2 A_a(\epsilon)+\Gamma_b A_b(\epsilon)]
\nonumber \\
&-\frac{1}{2\pi \hbar}\fint d\epsilon [f(\epsilon-eV^s)-f(\epsilon+eV^s)]\frac{2(\epsilon^2+\Gamma_a \Gamma_b +(\mu_b g_i H)^2)\left(\frac{\Gamma_L-\Gamma_R}{2}\right)^2}{|(\epsilon+i\Gamma_a)(\epsilon+i\Gamma_b)-(\mu_B g_i H)^2|^2},\nonumber
\\
\label{Is}
\end{align}
where $A_b$ is defined as
\begin{equation}
A_b(\epsilon)=-{\rm Im}\left\{\frac{\epsilon+i\Gamma_a}{(\epsilon+i\Gamma_a)(\epsilon+i\Gamma_b)-(\mu_B g_i H)^2} \right\}.
\end{equation}
We can immediately notice that eq.(\ref{Is}) reproduces eq.(6.7) in ref. \ref{Schiller2} if we set $V^s=0$. We also notice that we have finite spin current $I_s$ even though the case where $J^{LL}_{\perp}=J^{RR}_{\perp}$, i.e., {\it symmetric} case due to the existence of $\Gamma_b$.\\
Let us now discuss the spin current for specific cases, (i)symmetric, zero magnetic field, at finite temperature and (ii)symmetric, finite magnetic field, at $T=0$ in full detail. If we set $J^{LL}_{\perp}=J^{RR}_{\perp}$, $\Gamma_L=\Gamma_R$ and $\Gamma_2=0$, then we can write down $I_s$ as
\begin{equation}
I_s(V^s) = \frac{\Gamma_b}{2\pi \hbar} \fint d\epsilon A_b(\epsilon)[f(\epsilon-eV^ss)-f(\epsilon+eV^s)].
\label{Is0}
\end{equation}
Contrast to the charge current case (\ref{Ic}), $I_s$ involves $A_b$, i.e., the spectral function for the Majorana fermion $\hat b$ .
We can obtain a closed form expression for (\ref{Is0}) using the digamma function $\psi(z)$ \cite{Abr} which is defined by the gamma function $\Gamma(z)$ as 
\begin{equation}
\psi(z) = \frac{d}{dz} {\rm ln}\Gamma(z).
\end{equation}
For example, case (i), the spin current is rewritten as
\begin{equation}
I_s = \frac{\Gamma_b}{\pi \hbar}{\rm Im}\left\{\psi\left(\frac{1}{2}+\frac{\Gamma_b+ieV^s}{2\pi k_B T} \right) \right\},
\end{equation}
where $k_B$ is Boltzmann constant. For finite magnetic field case, we can also explicitly rewrite $I_s$ using the digamma function, however it takes a more complicated form. \\
\begin{figure}
\includegraphics[width=14cm,clip]{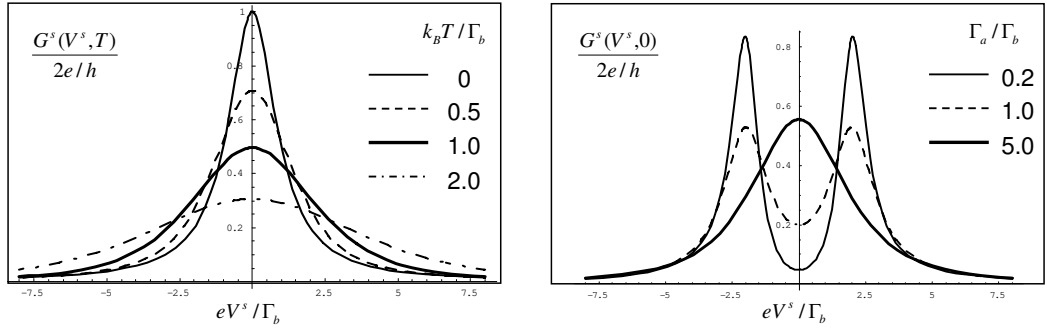}
\caption{The spin differential conductance $G_s(V^s,T)$ as a function of V 
(left) for zero magnetic field and different temperatures and (right) at 
$T=0$ and $\mu_B g_i H=2\Gamma_a$, for different ratios of $\Gamma_a$ to $\Gamma_b$.}
\label{fig1}
\end{figure}
Next, we will show the differential conductance for the spin current $G_s(V_s,T)\equiv dI_s/dV^s$ as a function of bias, for (i) and (ii). For (i), the result is
\begin{equation}
G_s(V^s,T)=\frac{2e}{h} \frac{\Gamma_b}{2\pi k_B T} {\rm Re}\left\{\psi^{(1)}\left(\frac{1}{2}+\frac{\Gamma_b+ieV^s}{2\pi k_B T} \right)\right\},
\end{equation}
where $\psi^{(1)}(z)\equiv d\psi(z)/dz$ denotes the trigamma function.
In the left figure of Fig.(\ref{fig1}), we show $G_s(V_s,T)$ for different temperatures. In the case (ii), we can easily carry out the integral, since the Fermi distribution function becomes the step function at zero temperature. Hence, the result is given by
\begin{equation}
G_s(V^s,0)=-\frac{2e}{h}\Gamma_b {\rm Im}\left(\frac{eV^s+i\Gamma_a}{(eV^s+i\Gamma_a)(eV^s+i\Gamma_b)-(\mu_B g_i H)^2} \right). 
\end{equation}
In the right figure of Fig.(\ref{fig1}), we show $G_s(V^s,0)$ for a moderately large magnetic field $\mu_B g_i H=2\Gamma_b$ at different ratios of $\Gamma_a$ to $\Gamma_b$. For $\Gamma_a=\Gamma_b$, magnetic field splits the zero-field resonance into two Lorentzians centered about $\pm \mu_B g_i H$.
\\
\\
\noindent
{\it Impurity Magnetization}\\
Next, we shall briefly discuss the impurity magnetization which provides direct information about the onset of Kondo screening. The average of the impurity magnetization $\la M^z \ra=\la i\mu_B g_i \hat a \hat b \ra$ can be computed as a function of the magnetic field $H$, the voltage $V^e$ and the spin voltage $V^s$. The resultant expression is
\begin{align}
&M(H,V^e,V^s)
\nonumber \\
&= \mu g_i \Bigg[ \frac{\Gamma_1}{\Gamma_a +\Gamma_b} \left(-\frac{\mu_B g_i H}{\pi}\fint d\epsilon \frac{(\Gamma_a+\Gamma_b)\epsilon}{|(\epsilon+i\Gamma_a)(\epsilon+i\Gamma_b)-(\mu_B g_i H)^2|^2} \right)f(\epsilon-eV^e)
\nonumber \\
&+ \frac{\Gamma_a+\Gamma_b-\Gamma_1}{\Gamma_a+\Gamma_b}\left(-\frac{\mu_B g_i H}{\pi}\fint d\epsilon \frac{(\Gamma_a+\Gamma_b)\epsilon}{|(\epsilon+i\Gamma_a)(\epsilon+i\Gamma_b)-(\mu_B g_i H)^2|^2}  \right)f(\epsilon-eV^s)
\nonumber \\
&+\frac{1}{2\pi}\frac{\Gamma_L-\Gamma_R}{4}\fint d\epsilon \frac{\epsilon^2+\Gamma_a \Gamma_b +(\mu_B g_i H)^2}{|(\epsilon+i\Gamma_a)(\epsilon+i\Gamma_b)-(\mu_B g_i H)^2|^2}[f(\epsilon-eV^s)-f(\epsilon+eV^s)].
\nonumber \\
\label{mag}
\end{align}
Setting $V^s=0$, eq.(\ref{mag}) reproduces eq.(8.2) in ref. \ref{Schiller2}.
We can also extract more explicit expression from eq.(\ref{mag}) using the digamma function $\psi(z)$. 
%(see Appendix ?).

\section{Conclusion}
In conclusion, we have exactly solved the nonequilibrium Kondo problem at the special point in the parameter space of the model with the spin-dependent chemical potentials. Using this solution, we have computed several experimentally observable quantities: charge current, spin current and magnetic properties. We have found that the spin current does not vanish even in the case  $J_{\perp}^{LL}=J_{\perp}^{RR}$, i.e., perpendicular part of the exchange interactions between two leads and spin are symmetric. Moreover, we have explicitly calculated the observables for the specific case in which the impurity spin is placed between two ferromagnetically polarized leads. 
This case, a common situation in spin transport phenomena, has never exactly been discussed out of equilibrium and under the Kondo temperature before. 
In the future, it will be interesting to study the nonequilibrium Kondo effect in a quantum dot coupled to two noncollinear ferromagnetic leads \cite{Matsubayashi}.

\begin{acknowledgments}
The author is grateful to K. Saito, A. Nishino, S. Murakami and N. Nagaosa for fruitful discussions. 
This work was supported by supported by Grant-in-Aids (Grant No. 15104006, No. 16076205, and No. 17105002) and NAREGI Nanoscience Project from the Ministry of Education, Culture, Sports, Science, and Technology.
HK is supported by the Japan Society for the Promotion of Science.
\end{acknowledgments}

%Chapter5


\begin{thebibliography}{200}
\bibitem{Kondo}
J. Kondo, Prog. Theor. Phys. {\bf 32} (1964) 37.
\bibitem{Ab}
A. Abrikosov, Physics {\bf 2} (1965) 5.
\bibitem{Anderson}
P. W. Anderson, J. Phys. C {\bf 3} (1970) 2436.
\bibitem{Yamada}
K. Yamada, Prog. Theor. Phys. {\bf 53} (1975) 970.
\bibitem{Tarucha}
S. Sasaki and S. Tarucha, J. Phys. Soc. Jpn. {\bf 74} (2005) 88.
\bibitem{Oguri}
A. Oguri, J. Phys. Soc. Jpn. {\bf 74} (2005) 110.
\label{Oguri}
\bibitem{Hewson}
A. C. Hewson: {\it The Kondo Problem to Heavy Fermions}(Canbridge University Press, Cambridge, 1997).
\label{Hewson}
\bibitem{Shiba}
{\it Kondo Effect-40 Years after the Discovery}, J. Phys. Soc. Jpn. {\bf 74}
\label{Shiba}
\bibitem{Wilson}
K. Wilson, Rev. Mod. Phys. {\bf 47} (1975) 773.
\bibitem{Andrei}
N. Andrei, K. Furuya, and J. H. Lowenstein, Rev. Mod. Phys. {\bf 55} (1983) 331.
\bibitem{Wiegmann}
A. M. Tsvelik and P. B. Wiegmann, Ad. Phys. {\bf 32} (1983) 453.
\bibitem{Wiegmann2}
P. B. Wiegmann, Phys. Lett. A {\bf 31} (1981) 392.
\bibitem{Affleck}
I. Affleck, Nucl. Phys. B {\bf 336} (1990) 517;
I. Affleck and A. W. W. Ludwig, Nucl. Phys. B {\bf 352} (1991) 849;
I. Affleck and A. W. W. Ludwig, Nucl. Phys. B {\bf 360} (1991) 649.
\bibitem{Toulouse}
G. Toulouse, Phys. Rev. B {\bf 2} (1970) 270.
\bibitem{Emery}
V. J. Emery and S. Kivelson, Phys. Rev. B {\bf 46} (1992) 10812.
\bibitem{Schiller}
A. Schiller and S. Hershfield, Phys. Rev. B {\bf 51} (1995) 12896;
\label{Schiller}
\bibitem{Schiller2}
A. Schiller and S. Hershfield, Phys. Rev. B {\bf 58} (1998) 14978.
\label{Schiller2}
\bibitem{Hershfield}
S. Hershfield, Phys. Rev. Lett. {\bf 70} (1992) 2134.
\label{Hershfield}
\bibitem{Delft}
J. Delft, A. W. W. Ludwig and V. Ambegaokar, Ann. Phys. {\bf 273} (1999) 175.
\bibitem{Mehta}
P. Mehta and N. Andrei, Phys. Rev. Lett {\bf 96} (2006) 216802.
\bibitem{Tasaki}
S. Tasaki, Chaos, Soliton and Fractals {\bf 12} (2001) 2657.
\bibitem{Takahashi}
J. Takahashi and S. Tasaki, J. Phys. Soc. Jpn, {\bf 74} (2005) Suppl., p. 261.
\bibitem{Furusaki}
A. Furusaki and N. Nagaosa, Phys. Rev. Lett. {\bf 72} (1994) 892.
\bibitem{AL}
I. Affleck and A. W. W. Ludwig, J. Phys. A {\bf 27} (1994) 5375.
\label{AL}
\bibitem{Abr}
{\it Handbook of Mathematical Functions}, ed. M. Abramowitz and I. A. Stegun (Dover, New York, 1972) p. 79.
\bibitem{Matsubayashi}
D. Matsubayashi and M. Eto: cond-mat/0607548.
\end{thebibliography}
\end{document}